\documentstyle[12pt]{article}

\catcode`\@=11
\long\def\@makefntext#1{
\protect\noindent \hbox to 3.2pt {\hskip-.9pt
$^{{\ninerm\@thefnmark}}$\hfil}#1\hfill}		%CAN BE USED

\def\@makefnmark{\hbox to 0pt{$^{\@thefnmark}$\hss}}  %ORIGINAL

\def\ps@myheadings{\let\@mkboth\@gobbletwo
\def\@oddhead{\hbox{}
\rightmark\hfil\ninerm\thepage}
\def\@oddfoot{}\def\@evenhead{\ninerm\thepage\hfil
\leftmark\hbox{}}\def\@evenfoot{}
\def\sectionmark##1{}\def\subsectionmark##1{}}

\renewcommand{\thefootnote}{\fnsymbol{footnote}}

\newcounter{sectionc}\newcounter{subsectionc}\newcounter{subsubsectionc}
\renewcommand{\section}[1] {\vspace*{0.6cm}\addtocounter{sectionc}{1}
\setcounter{subsectionc}{0}\setcounter{subsubsectionc}{0}\noindent
	{\normalsize\bf\thesectionc. #1}\par\vspace*{0.4cm}}
\renewcommand{\subsection}[1] {\vspace*{0.6cm}\addtocounter{subsectionc}{1}
	\setcounter{subsubsectionc}{0}\noindent
	{\normalsize\it\thesectionc.\thesubsectionc. #1}\par\vspace*{0.4cm}}
\renewcommand{\subsubsection}[1]
{\vspace*{0.6cm}\addtocounter{subsubsectionc}{1}
	\noindent {\normalsize\rm\thesectionc.\thesubsectionc.\thesubsubsectionc.
	#1}\par\vspace*{0.4cm}}

\newcounter{appendixc}
\newcounter{subappendixc}[appendixc]
\newcounter{subsubappendixc}[subappendixc]

\renewcommand{\appendix}[1] {\vspace*{0.6cm}
        \refstepcounter{appendixc}
        \setcounter{figure}{0}
        \setcounter{table}{0}
        \setcounter{equation}{0}
        \renewcommand{\thefigure}{\Alph{appendixc}.\arabic{figure}}
        \renewcommand{\thetable}{\Alph{appendixc}.\arabic{table}}
        \renewcommand{\theappendixc}{\Alph{appendixc}}
        \renewcommand{\theequation}{\Alph{appendixc}.\arabic{equation}}
        \noindent{\bf Appendix \theappendixc #1}\par\vspace*{0.4cm}}

\def\abstracts#1{{

\centering{\begin{minipage}{12.2truecm}\footnotesize\baselineskip=12pt\noindent
	\centerline{\footnotesize ABSTRACT}\vspace*{0.3cm}
	\parindent=0pt #1
	\end{minipage}}\par}}
\renewenvironment{thebibliography}[1]
	{\begin{list}{\arabic{enumi}.}
	{\usecounter{enumi}\setlength{\parsep}{0pt}
\setlength{\leftmargin 1.25cm}{\rightmargin 0pt}
	 \setlength{\itemsep}{0pt} \settowidth
	{\labelwidth}{#1.}\sloppy}}{\end{list}}

\newcounter{itemlistc}
\newcounter{romanlistc}
\newcounter{alphlistc}
\newcounter{arabiclistc}

\newcommand{\fcaption}[1]{
        \refstepcounter{figure}
        \setbox\@tempboxa = \hbox{\footnotesize Fig.~\thefigure. #1}
        \ifdim \wd\@tempboxa > 6in
           {\begin{center}
        \parbox{6in}{\footnotesize\baselineskip=12pt Fig.~\thefigure. #1}
            \end{center}}
        \else
             {\begin{center}
             {\footnotesize Fig.~\thefigure. #1}
              \end{center}}
        \fi}

\newcommand{\tcaption}[1]{
        \refstepcounter{table}
        \setbox\@tempboxa = \hbox{\footnotesize Table~\thetable. #1}
        \ifdim \wd\@tempboxa > 6in
           {\begin{center}
        \parbox{6in}{\footnotesize\baselineskip=12pt Table~\thetable. #1}
            \end{center}}
        \else
             {\begin{center}
             {\footnotesize Table~\thetable. #1}
              \end{center}}
        \fi}

\def\@citex[#1]#2{\if@filesw\immediate\write\@auxout
	{\string\citation{#2}}\fi
\def\@citea{}\@cite{\@for\@citeb:=#2\do
	{\@citea\def\@citea{,}\@ifundefined
	{b@\@citeb}{{\bf ?}\@warning
	{Citation `\@citeb' on page \thepage \space undefined}}
	{\csname b@\@citeb\endcsname}}}{#1}}

\newif\if@cghi
\def\cite{\@cghitrue\@ifnextchar [{\@tempswatrue
	\@citex}{\@tempswafalse\@citex[]}}
\def\citelow{\@cghifalse\@ifnextchar [{\@tempswatrue
	\@citex}{\@tempswafalse\@citex[]}}
\def\@cite#1#2{{$\null^{#1}$\if@tempswa\typeout
	{IJCGA warning: optional citation argument
	ignored: `#2'} \fi}}

 1
\font\twelverm=cmr10  scaled\magstep 1
 1

\font\tenrm=cmr10

\font\ninerm=cmr9

\begin{document}

\newcommand{\st}{\scriptstyle}
\newcommand{\sst}{\scriptscriptstyle}
\newcommand{\mco}{\multicolumn}
\newcommand{\epp}{\epsilon^{\prime}}
\newcommand{\vep}{\varepsilon}
\newcommand{\ppg}{\pi^+\pi^-\gamma}
\newcommand{\vp}{{\bf p}}
\newcommand{\ko}{K^0}
\newcommand{\kb}{\bar{K^0}}
\newcommand{\al}{\alpha}
\newcommand{\ab}{\bar{\alpha}}
\def\z2{\ifmmode Z_2\else $Z_2$\fi}
\def\ie{{\it i.e.},}
\def\eg{{\it e.g.},}
\def\etc{{\it etc}}
\def\etal{{\it et al.}}
\def\ibid{{\it ibid}.}
\def\gev{\,{\rm GeV}}
\def\to{\rightarrow}
\def\epem{\ifmmode e^+e^-\else $e^+e^-$\fi}
\def\Re{{\cal R \mskip-4mu \lower.1ex \hbox{\it e}\,}}
\def\Im{{\cal I \mskip-5mu \lower.1ex \hbox{\it m}\,}}
\def\be{\begin{equation}}
\def\ee{\end{equation}}
\def\bea{\begin{eqnarray}}
\def\eea{\end{eqnarray}}
\def\CPbar{\hbox{{\rm CP}\hskip-1.80em{/}}}%temp replacement due to no font

\rightline{\vbox{\halign{&#\hfil\cr
&SLAC-PUB-95-6741\cr}}}

\centerline{{\normalsize\bf EXTENDED GAUGE SECTORS}
\footnote{Work supported by the Department of
Energy, contract DE-AC03-76SF00515.}
\footnote{Presented at the {\it Fourth International Conference on
Physics Beyond the Standard Model}, Lake Tahoe, CA, December 13-18, 1994.}
}
\baselineskip=15pt

%\vfill
%\vspace*{0.6cm}
\centerline{\footnotesize THOMAS G. RIZZO}
\baselineskip=13pt
\centerline{\footnotesize\it Stanford Linear Accelerator Center,
Stanford University}
\baselineskip=12pt
\centerline{\footnotesize\it Stanford, CA 94309, USA}
\centerline{\footnotesize E-mail: rizzo@slacvx.slac.stanford.edu}

%\vfill
\vspace*{0.9cm}
\abstracts{Present and future prospects for the discovery of new gauge bosons,
$Z'$ and $W'$, are reviewed. Particular attention is paid to hadron and
$e^+e^-$ collider searches for the $W'$ of the Left-Right Symmetric Model.}

%\vspace*{0.6cm}
\normalsize\baselineskip=15pt
\setcounter{footnote}{0}
\renewcommand{\thefootnote}{\alph{footnote}}
\section{Introduction}
An extension of the gauge sector of the Standard Model(SM) would not only lead
to the existence of new gauge fields, but will almost always require
the introduction of
exotic fermions{\cite {djouadi}} to cancel anomalies as well as new Higgs
fields{\cite {haber}} to break the
extended gauge symmetry. In addition, GUT scenarios leading to gauge
extensions require the existence of SUSY in order to maintain the hierarchy
of breaking scales
and obtain coupling constant unification. Thus the phenomenology of extended
gauge models(EGM) is particularly rich as is indicated by the rather
extensive literature on
this subject. Unfortunately, this implies that there are an enormous
number of interesting models currently on the market which means that any
overview of the
subject is necessarily incomplete. Hence, we will be forced to
limit ourselves to
a few representative models and restrict our discussion to searches for
new gauge bosons at hadron and $e^+e^-$ colliders{\cite {mk}}. Regrettably,
this leaves vast and fascinating territories untouched.

In what
follows, we chose as examples the set of models recently discussed by
Godfrey{\cite {steve}} so that we need say little here about the coupling
structure of each scenario; curious readers are requested to consult
Godfrey's paper and references therein for the details of each model. To
be specific, we
consider ({\it i}) the $E_6$ effective rank-5 model(ER5M), which predicts a
$Z'$ whose couplings depend on a single parameter
$-\pi/2 \leq \theta \leq \pi/2$ (with models $\psi$, $\chi$, and $\eta$
denoting specific $\theta$ values); ({\it ii}) the Sequential Standard
Model(SSM)
wherein the new $W'$ and $Z'$ are just heavy versions of the SM particles (of
course, this is not a true model in the strict sense but is commonly used as a
guide by experimenters); ({\it iii}) the Left-Right Symmetric Model(LRM)
and, lastly, ({\it iv}) the Alternative Left-Right Model(ALRM), arising from
$E_6$, wherein the fermion assignments are modified in comparison to the LRM.
In the ALRM, the $W'$ carries lepton number so that it cannot be produced via
the ordinary Drell-Yan process but only in association with a leptoquark thus
making it difficult to observe over top quark backgrounds at hadron colliders.
The LRM owes much of its survival over the last two decades to the plethora
of free parameters it contains: ({\it a}) the ratio of the gauge couplings,
$0.55 \leq \kappa=g_R/g_L \leq 2(naturalness??)$, the lower limit being forced
upon us by the internal consistency of the model;
({\it b}) the masses of the
right-handed(RH) neutrinos, ({\it c}) the elements of the RH CKM mixing
matrix, $V_R$, which are {\it a priori} different than $V_L$, and ({\it d}) the
$W_R$-$Z_R$ mass relationship,
\begin {equation}
{M_{W_R}^2\over {M_{Z_R}^2}} =  {{(1-x_w)\kappa^2-x_w}\over {\rho_R(1-x_w)
\kappa^2}}
\end {equation}
where $x_w$ is the usual weak mixing angle and the parameter $\rho_R$ takes
on the value 1(2) if the $SU(2)_R$
breaking sector consists solely of Higgs doublets(triplets). (The triplet
scheme is favored in the see-saw scenario for neutrino masses.) From this we
see that unless the $SU(2)_R$ breaking sector is somewhat unusual, the
$Z_R$ will always be more massive that the $W_R$. This large set of parameters
will return to haunt us when we examine $W_R$ searches.

\section{$Z'$ : Then and Now}
Since $Z'$ searches have been discussed by many authors{\cite {steve}}, our
overview of this subject will be quite brief.
At present, the Tevatron provides the best direct search limits for new gauge
bosons{\cite {cdf}}, corresponding to 505 GeV for the $Z'$ (and 652 GeV for the
$W'$) of the SSM, from the run Ia electron data sample. Figs.1a-c
show how the $Z'$ search reach of the Tevatron should evolve with time for
several different models assuming no new particles are discovered;
including $\mu$'s in the data sample should increase all of the results shown
by $\simeq 35-40$ GeV. In all cases, we assume that the $Z'$ decays to only SM
fermions and $Z$-$Z'$ mixing is neglected. Apart from these assumptions, the
limits depend only upon a single parameter, $\theta$ in the ER5M and $\kappa$
in the LRM. Pushing the Tevatron luminosity, {\it L}, up
above 1 $fb^{-1}$ implies that $Z'$ masses of order 1 TeV are beginning to be
probed. Figs.1d-f show the corresponding (electrons {\it only}!)
results for the LHC(with $\sqrt {s}=$14 TeV) and the influence of additional
decay modes on the
search reach, \ie ~decreasing the leptonic branching fraction of the $Z'$ by a
factor of 2 reduces the reach by $\simeq 0.33$ TeV. For LHC luminosities above
100 $fb^{-1}$, $Z'$ masses in excess of 4 TeV become accessible. At the NLC,
$Z'$ searches are performed by looking for systematic shifts in multiple
observables, making full use of the anticipated high electron
beam polarization. A 500
GeV machine with {\it L}=50 $fb^{-1}$ probes $Z'$ masses in the 1.5-5 TeV
range{\cite {steve}}, which nicely complements the direct production searches
at the LHC. A machine with four times this energy and luminosity may
extend this reach by a factor of 3-4.

\section{$W'$ : Hadron Collider Search Caveats}
Unlike $Z'$ searches at hadron colliders, the corresponding $W'$ searches
via the Drell-Yan process have
many subtleties; this is most easily demonstrated within the LRM
context{\cite {me1}}. The
CDF $W'$ search assumes that the $q'\bar qW'$ production vertex has SM
strength (\ie ~({\it i}) $\kappa=1$ and
({\it ii}) $|V_{L_{ij}}|=|V_{R_{ij}}|$),
that the RH neutrino is ({\it iii}) `light' and `stable', appearing as missing
$E_T$ in the detector, and that the $W_R$ leptonic branching
fraction($B_l$) is the SM value apart from contributions due to open
top(\ie ~({\it iv}) no exotic decay channels are
open). If any of these assumptions are invalid, what happens to the
search reach? Assumptions
({\it i}) and ({\it iv}) are easily accounted for by the introduction of an
effective $\kappa$ parameter, $\kappa_{eff}=\kappa \sqrt{B_l/B_l^{SSM}}$
which simply adjusts the overall cross-section normalization with
the resulting reach shown in Fig.2a. If assumption ({\it ii}) is invalid, a
significant search reach degradation{\cite {me1}} occurs as is shown in
Fig.2b for CDF run Ia; \eg ~one finds via a Monte Carlo study that for
$50(10)\%$ of the $V_R$ parameter space the Tevatron run Ia
$W_R$ reach is reduced to less than 550(400) GeV. This reduction is a result
of modifying the weight of the various parton luminosities which enter into
the calculation of the cross-section. At the LHC, surrendering ({\it ii}) does
not cost us such a large penalty since the $W_R$ production process occurs
through the annihilation of sea$\times$valence quarks in $pp$ collisions,
whereas it is a valence$\times$valence process at the Tevatron. From
Fig.2c we see that varying $V_R$ modifies the reach no more than $20\%$. Life
gets {\it much} harder if $\nu_R$ does not appear as missing $E_T$. A massive
$\nu_R$ will most likely decay within the detector to ${\ell}^{\pm}+jj$, with
either charge sign equally likely if $\nu_R$ is a Majorana fermion. A parton
level analysis of this scenario has been carried out by
Datta \etal {\cite {roy}} for the LHC; they find a `viable signal' for $W_R$
masses below 2-3 TeV for the entire $m_{\nu_R}<M_{W_R}$ range. (This analysis
needs
to be repeated including a full detector simulation and should also be done
for the Tevatron.) Perhaps the worst case scenario is when $\nu_R$ is more
massive than $W_R$ so that $W_R$ has only hadronic (or exotic) decay channels
open. Can $W_R$ be seen as a bump in dijets? Clearly the chances are somewhat
better at the Tevatron where $S/B$ is perhaps manageable given reasonable
statistics; CDF has already performed such an analysis with run Ia
data{\cite {dijet}} with somewhat limited results. At the LHC, where the dijet
backgrounds have increased enormously due to the rise in the glue-glue
luminosity, a preliminary study by the ATLAS Collaboration indicates that such
dijet searches might still be possible provided excellent energy resolution is
available{\cite {dijet}}. More analysis is necessary to clarify this case.

Additional help in such a pessimistic situation may be provided by the LRM's
$W_R$-$Z_R$ mass relationship, \ie  ~if a $Z_R$ is found but
$m_{\nu_R}>M_{W_R}$,
this relation tells us something about {\it where} to look in dijets for the
$W_R$. If, instead, only a limit on the $Z_R$ mass is obtained, the same mass
relation can be used to get a relatively weak (but conservative!) limit
on the mass of $W_R$. Figs.2e-f
show the result of this approach for the Tevatron using the curves in Fig.1b
as input. Note the indirect limit on the $W_R$ mass from run Ia with
$\kappa=1$ is only 270 GeV assuming triplet $SU(2)_R$ breaking, which is only
about $45\%$
of the canonical SSM value. When the integrated {\it L} increases to 1
$fb^{-1}$, this bound grows to only 450 GeV.
This indirect limit is substantially larger at
the LHC, as shown in Fig.2f, but is still less than $50\%$ of the usually
claimed reach. Note that this limit is reasonably sensitive to the nature of
$SU(2)_R$ breaking but somewhat less sensitive as to whether the $Z_R$ has
exotic decay modes. If dijet $W_R$ searches are impossible in practice, we
need to turn to other production strategies.

\section{$W_R$'s at the NLC}
The NLC can also play a crucial role at unraveling the charged-current
sector of EGM's.
$W_R$ production in $e^+e^-$, $\gamma e$, and $e^-e^-$ collisions
{\cite {eeww}} is insensitive to $V_R$ and scales
simply with $\kappa$ thus immediately avoiding two of the above difficulties
with hadron collider searches. All three processes can yield valuable
information about both $W_R$ and the mass spectrum of the LRM.
Note that the like-sign $e^-e^-$
process {\it only} occurs when $\nu_R$ is a Majorana fermion.
In addition, due to the relatively clean
environment and high beam polarization, signatures are also easier to spot and
backgrounds are readily reduced.
Unfortunately, the sensitivity to $m_{\nu_R}(=M_N)$ remains at some level in
all cases and a dependence on the doubly-charged Higgs mass, $M_\Delta$,
occurs in the $e^-e^-$ case.

$W_R$ pair production occurs with a large $\sigma$ yielding
more that $10^{4}$ events up to the
kinematic limit as shown in Figs.3a-b; increasing the $\nu_R$ mass in
the $t$-channel graph generally reduces $\sigma$ near threshold, where $\sigma$
is largest, and flattens the angular distribution. For large $\sqrt {s}$ it
delays the unitarity cancellation between the amplitudes resulting in a bigger
$\sigma$. Since the $Z_R$ mass is less than twice that of $W_R$ for most
parameter values, $\sigma$ does not show much sensitivity to the possible
variations in $M_{Z_R}$.
For reasonable {\it L}'s, $W_R(W_R)^*$ production allows for searches
up to $M_R\simeq 0.8\sqrt {s}$. At the tree level, the $W_R$ pair
cross-section is insensitive to the Dirac or Majorana nature of the RH
neutrino.

The single production of $W_R$'s in association with $\nu_R$ in $\gamma e$
collisions via laser backscattering has been re-analyzed recently
by Raidal{\cite {eeww}} taking into account both $e$ and $\gamma$
beam polarization. Essentially the entire kinematic region is found to be
accessible with polarization playing an important role in identifying the
signal and reducing backgrounds.

The $e^-e^-\to W_R^-W_R^-$ lepton-number violating process is perhaps
the most interesting way of looking for $W_R$'s as both
the Majorana nature of $\nu_R(N)$ and the $SU(2)_R$ symmetry breaking are
probed
simultaneously. The helicity-amplitude analysis for like-sign production
has recently been performed by Helde \etal{\cite {eeww}}. As shown there, as
well as in previous analyses(see Figs.3c-d), the cross-sections are quite
large but reasonably sensitive to both $M_{N,\Delta}$ variations. As a whole,
larger values of $M_N$ yield larger rates whereas the cross-section vanishes
as $M_N\to 0$. It has recently been shown that
allowing for one of the $W_R$'s to be off-shell still yields a reasonable
rate for $W_R$ masses as large as $0.8\sqrt {s}$(see Figs.3e-f). This analysis
assumed that only the {\it {jj}} decay modes of the $W_R$ were accessible
thus allowing for the possibility of $M_N>M_R$. In either case, the $W_R$
angular distribution is found to be relatively flat implying that acceptance
cuts will not have any substantial impact on rates.

\section{References}

%%%%%%%%%%%%%%%%%%%%%%%%%%%%%%%%%%%%%%%%%%%%%%%%%%%%%%%
\def\MPL #1 #2 #3 {Mod.~Phys.~Lett.~{\bf#1},\ #2 (#3)}
\def\NPB #1 #2 #3 {Nucl.~Phys.~{\bf#1},\ #2 (#3)}
\def\PLB #1 #2 #3 {Phys.~Lett.~{\bf#1},\ #2 (#3)}
\def\PR #1 #2 #3 {Phys.~Rep.~{\bf#1},\ #2 (#3)}
\def\PRD #1 #2 #3 {Phys.~Rev.~{\bf#1},\ #2 (#3)}
\def\PRL #1 #2 #3 {Phys.~Rev.~Lett.~{\bf#1},\ #2 (#3)}
\def\RMP #1 #2 #3 {Rev.~Mod.~Phys.~{\bf#1},\ #2 (#3)}
\def\ZP #1 #2 #3 {Z.~Phys.~{\bf#1},\ #2 (#3)}
\def\IJMP #1 #2 #3 {Int.~J.~Mod.~Phys.~{\bf#1},\ #2 (#3)}

\bibliographystyle{unsrt}

\newpage
\tenrm{

%$   figure  captions   these need to be put in 10pt !!!!!
{%\small
%\vspace*{2.00in}
\noindent
Fig.~1. Tevatron search reach for the $Z'$ in the (a)ER5M and (b)LRM for run
Ia(lower curves, MRSA pdf's are dashdots while CTEQ3M pdf's are solid) and
with increased {\it L}'s of 100, 250, 500, and 1000
$pb^{-1}$(from bottom to top). (c){\it L} dependence of Tevatron search reach
for the ALRM(dashdot), SSM(dots), LRM with $\kappa=1$(dashes), and
$\psi$(solid) $Z'$'s. (d) and (e) are the same as (a) and (b) but for the LHC
with 100 $fb^{-1}$; the lower curve corresponds to a reduction of the naive
leptonic branching fraction by a factor of 2. (f)Same as (c) but for the LHC.

\medskip

%\vspace*{8.15in}
\noindent
Fig.~2. (a)Tevatron $W_R$ reach as a function of $\kappa_{eff}$ as described in
the text for the same {\it L} values as in Fig.1a. (b)Percentage of the $V_R$
parameter space allowing the $W_R$ below a given value from run Ia.
(c)Maximum and minimum cross-sections for $W_R$ production at the LHC due to
$V_R$ variations for $\kappa=1$. Indirect $W_R$ search limits for the Tevatron
(d)run Ia and with (e){\it L}=1 $fb^{-1}$ as well as (f)for the LHC.
Doublet(triplet) $SU(2)_R$ breaking corresponds to the dotted(dashdotted)
curves. In (f), the lower curves correspond to a factor of 2 reduction in
the $Z'$ leptonic branching fraction.

\medskip

%\vspace*{8.15in}
\noindent
Fig.~3. (a)$W_R$ pair production cross-section vs. $M_N$ at a
1.5 TeV NLC assuming $\kappa$=1 and $M_R$=700 GeV. (b)Same as (a) but vs.
$\sqrt {s}$ assuming
$M_N$=100(500,1000,2000) GeV corresponding to the
dotted(dashed,dashdotted,solid) curve.
Cross-section for like-sign $W_R$ production with ${\sqrt {s}}$=1 TeV
as a function of (c)$M_N$
and (d)$M_{\Delta}$ for $\kappa$=0.9 and $M_R$=480 GeV. In[(c),(d)],
the curves on the right(left)-hand side correspond, from top to bottom, to
$M_{\Delta}$=800,1200,500,1500,200, and 2000 GeV [$M_N$=1500,1200,800,
500, 200 GeV].
Event rates per 100 $fb^{-1}$ for $W_R+jj$ production at a
1.5 TeV $e^-e^-$ collider assuming $\kappa=1$ and $M_R$=1 TeV
(e)as a function of $M_N$ for $M_{\Delta}$=0.3(0.6,1.2,1.5,2) TeV
corresponding to the dotted(dashed, dash-dotted, solid, square-dotted) curve;
(f)as a function of $M_{\Delta}$ for $M_N$=0.2(0.5,0.8,1.2,1.5) TeV
corresponding to the dotted(dashed, dash-dotted, solid, square-dotted) curve.

}
}
\twelverm

\end{document}